\UseRawInputEncoding

\documentclass[aps,prl,reprint,superscriptaddress,nobibnotes]{revtex4-2}

\usepackage{float}
\usepackage{hyperref}
\usepackage{xcolor}
\usepackage{soul}
\usepackage{graphicx}
\usepackage{svg}
\usepackage{yfonts}
\usepackage{dcolumn}
\usepackage{bm}

\usepackage{ulem}

\begin{document}

\preprint{APS/123-QED}

\title{\textbf{Vortex Tunneling and Critical State in an Oxide Heterostructure} 
}%


\author{Jordan T. McCourt}
\email{jordan.mccourt@duke.edu}
\affiliation{%
Department of Physics, Duke University, Durham, NC 27708, USA
}%

\author{Ryan Henderson}
\affiliation{%
Department of Physics, Duke University, Durham, NC 27708, USA
}%

\author{John Chiles}
\affiliation{%
Department of Physics, Duke University, Durham, NC 27708, USA
}%

\author{Chun-Chia Chen}
\affiliation{%
Department of Physics, Duke University, Durham, NC 27708, USA
}%

\author{Shama}
\affiliation{%
Department of Physics, Duke University, Durham, NC 27708, USA
}%

\author{Divine Kumah}
\affiliation{%
Department of Physics, Duke University, Durham, NC 27708, USA
}%

\author{Vadim Geshkenbein}
\affiliation{%
Institute for Theoretical Physics, ETH Zurich, Zurich, Switzerland
}%

\author{Gleb Finkelstein}
\email{gleb@duke.edu}
\affiliation{%
Department of Physics, Duke University, Durham, NC 27708, USA
}%

\date{\today}

\begin{abstract}
Two-dimensional superconductors offer an excellent platform for the study of vortex matter due to their low superfluid stiffness and inability to effectively screen applied magnetic fields. Here we explore vortices in a two-dimensional superconductor formed at the surface of the complex oxide KTaO$_3$. Multiple regimes of vortex-mediated transport are identified and studied, revealing switching behaviour attributed to nucleation of individual vortices. Analysis of this regime allows us to identify the quantum tunneling of vortices, which transitions to thermally activated behaviour at elevated temperatures. Magnetic field dependence reveals rich histograms of the switching currents which we attribute to different configurations of pinned vortices.

\end{abstract}

\maketitle

\section{\label{sec:intro}Introduction}

Two-dimensional superconductors (2D) formed at complex oxide interfaces and van der Waals materials have attracted intense interest because they realize regimes fundamentally different from those found in conventional superconducting thin films. In these systems, superconductivity emerges in a truly two-dimensional environment where carrier density, spin–orbit coupling, and interaction strength can be tuned in situ~\cite{balents_superconductivity_2020,hwang_emergent_2012, zubko_interface_2011}. This tunability, together with reduced dimensionality, enhances the role of phase fluctuations, broken symmetries, and electronic correlations, enabling access to superconducting states that are difficult or impossible to achieve in bulk materials. As a result, 2D superconductors provide a versatile platform for exploring unconventional pairing, quantum phase transitions, and the interplay between superconductivity, topology, and other competing electronic orders.

In 2D superconductors with thickness $d$ much less than the London penetration depth $\lambda$, the characteristic length determining the screening of magnetic field is the Pearl length $\Lambda=\frac{2\lambda^2}{d}$~\cite{pearl_current_1964}. This scale could be as large as millimeters, which greatly exceeds the typical sample width $W$, as a result, the superconductor is unable to screen the applied perpendicular magnetic field, and $B \approx H$.
In this regime, the effects of surface nucleation and bulk pinning of vortices will dominate the transport properties. This regime is discussed in the recent paper~\cite{gaggioli_superconductivity_2024}, which considers a current-carrying superconducting strip in magnetic field. 

At very small magnetic fields (Meissner state) the field-induced diamagnetic currents on the two sides of the strip are either added to or subtracted from the externally applied bias current. As a result, the critical current density is reached on one side of the strip at a lower bias current compared to zero field. The measured critical current of the strip thus decreases linearly with the field, $I_{C}(H) = I_C(0)\left(1-\frac{|H|}{H_S}\right)$. Here, $I_C(0)$ is the critical current at zero field, and $H_S$ is the maximal field at which the Meissner state could exist at zero applied current.

While at small magnetic fields the current density at $I_C$ is too high to allow for the vortex pinning in the bulk, above a characteristic field $H^*$, the critical current density drops sufficiently to allow for the vortices to be pinned. The superconductor enters a critical state in which vortices can be present in the bulk, and $I_C(H)$ is increased to $I_P\left(1+\frac{(H^*)^2}{|H|H_P}\right)$~\cite{gaggioli_superconductivity_2024}, (here, $I_P$ and $H_P$ are parameters that depend on the vortex pinning). The deviation from the linear $I_C$ vs $H$ dependence and the onset of the characteristic $1/H$ dependence therefore signals the presence of vortices close to $I_C$. 

In this paper, we explore the regimes where a small number of vortices, possibly zero or one, are present in the superconducting constriction. We demonstrate the regimes corresponding to continuous vortex flow, and switching controlled by nucleation of a single vortex. In the latter regime, we find the temperature-independent saturation of the nucleation rate, which we associate with the macroscopic quantum tunneling of the vortex. 

\begin{figure*}[htp!]
\includegraphics[width=0.9\textwidth]{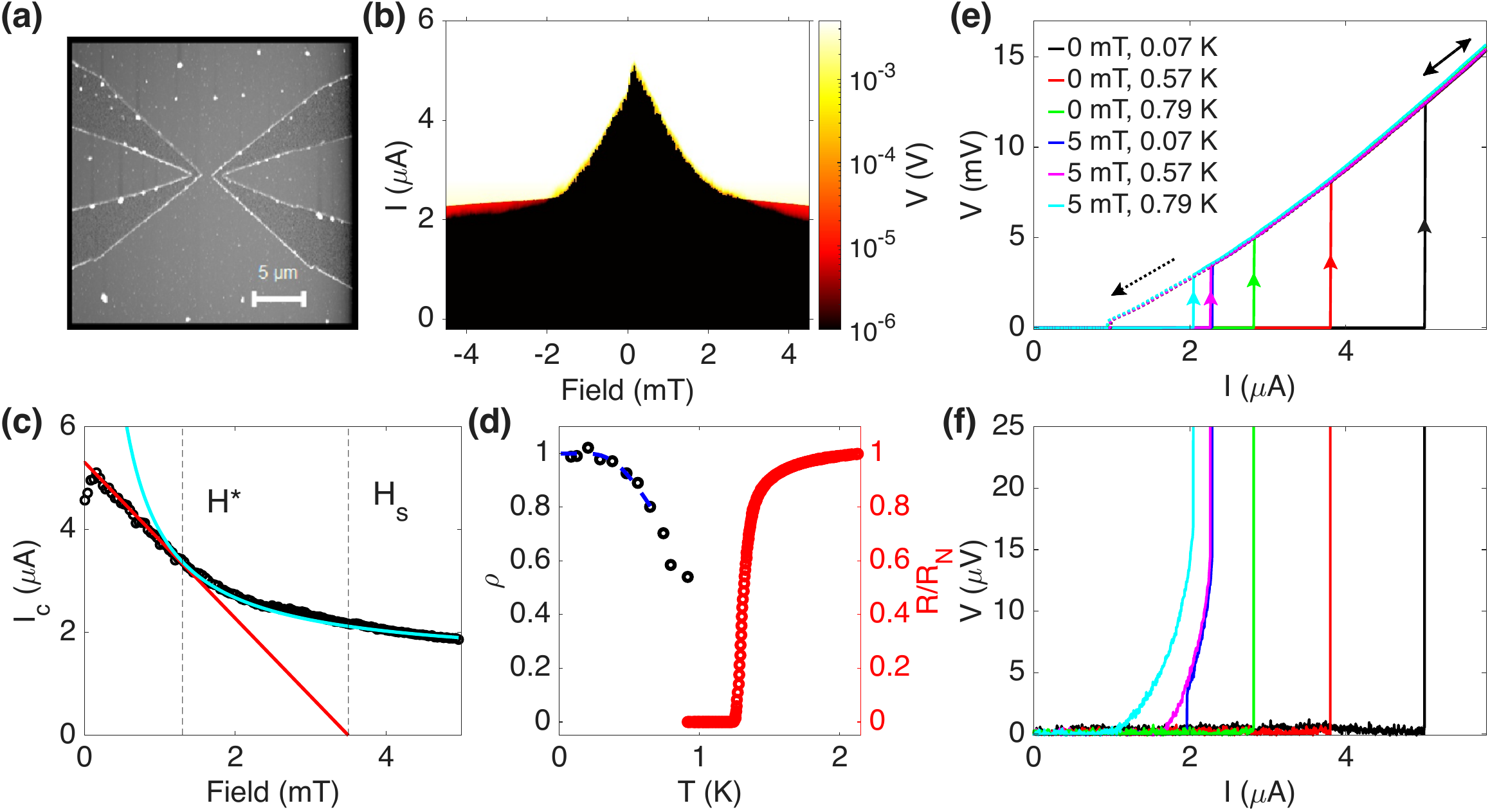}
\caption{\label{Lambda} \textbf{(a)} An AFM image of constriction A studied here and in Figure~\ref{IVs}. The light grey region denotes the conductive channel, whilst the darker regions denote the area etched and left insulating. The conductive regions around the constriction were intended to be used as gates in the style of our previous work~\cite{mccourt_electrostatic_2025}, but were left unused here. \textbf{(b)}~Map of voltage across the constriction as a function of bias current $I$ and perpendicular magnetic field $B$. Note the logarithmic colour scale. \textbf{(c)} Extracted $I_C$ (circles) overlayed with curves corresponding to expressions from \cite{gaggioli_superconductivity_2024}. Due to the small asymmetry of the junction (Supplementary Figure~\ref{supp_QPC3UpAndDown}) the max $I_C$ is shifted from zero field. \textbf{(d)} Relative superfluid density, $\rho$ (black circles) with fit (blue dotted line) and the  resistance of the film, $R(T)$, normalized to the normal state resistance, $R_N$ (red circles).   $\rho$ is obtained from the slope $dI_C/dB$ extracted in the Meissner regime at different temperatures (supplementary Figure~\ref{supp_IcIrVsT}). These slopes are then normalized to $dI_C/dB$ measured at the base temperature in panels (b) and (c). \textbf{(e)} IV curves for a range of temperatures at zero field and 5~mT. Solid and dashed lines correspond to IV sweeps with increasing and decreasing bias, respectively. Note that the switching currents depend on $T$ and $B$, while the retrapping currents and the $I-V$ curves in the high voltage state do not. These observations indicate that on the high voltage branch some part of the sample enters a normal state, which is then sustained by the dissipated power. \textbf{(f)} Same curves as in panel (e) on a much smaller voltage scale. The curves measured at finite field display an initial $\mu$V-scale segment, which we associate with the vortex flow.
}
\end{figure*}

\section{\label{sec:IcB}Critical current and field dependence}
\begin{figure*}[htp!]
\includegraphics[width=0.9\textwidth]{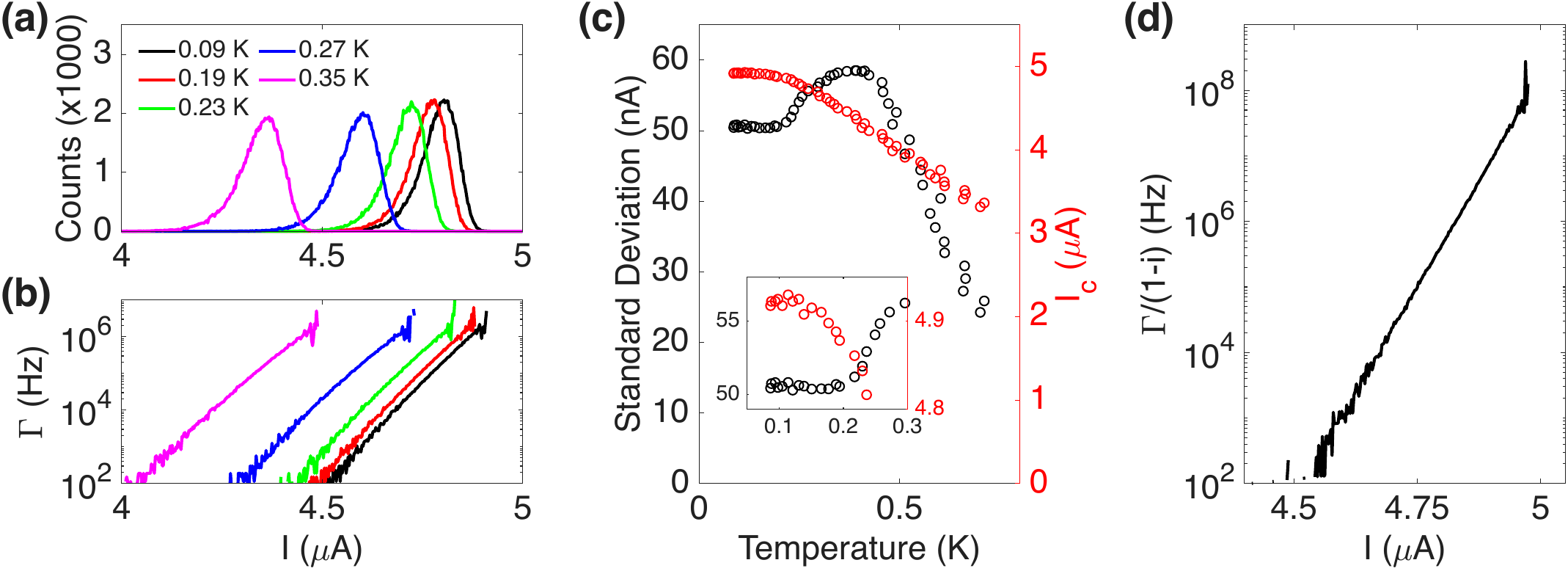}
\caption{\label{rate_T} \textbf{(a)} Histograms of switching currents at zero field and several temperatures for device B. \textbf{(b)} Escape rate extracted from the histograms in panel (a). \textbf{(c)} Standard deviation of the switching current distributions and critical current over a range of temperatures. \textbf{(d)} Scaling of the escape rate divided by the current-dependent prefactor (see text), demonstrating the exponential dependence on the applied current.}
\end{figure*}

The samples in this work were formed by depositing 7~nm of Al on a KTaO$_3$ (111) substrate. 
The Al layer is oxidized thereby reducing the KTaO$_3$ layer, as a result forming a 2-dimensional electron gas (2DEG) at the AlO$_x$/KTaO$_3$ interface~\cite{mallik_superfluid_2022}. The carrier density of the 2DEG is $n\approx8\times10^{13}$~cm$^{-2}$ with a mobility of $\approx90$~cm$^2$/Vs at cryogenic temperatures. The critical temperature is 1.3~K and the critical perpendicular magnetic field  0.6~T. Using electron beam lithography and reactive ion etching, we pattern 1~$\mu$m wide hourglass-shaped constrictions in the plane of the 2DEG, Figure~\ref{Lambda}a. Two devices are presented in the main text, refered to as devices A (Figures~\ref{Lambda},~\ref{IVs}) and B (Figures~\ref{rate_T},~\ref{rate_B}). 

Since the largest current density along the edge is concentrated at the constriction, we expect it to dominate the transition from the superconducting state. Figure~\ref{Lambda}b shows the voltage across the constriction, $V$, as a function of the applied current, $I$, and perpendicular magnetic field, $B$. We average multiple sweeps of $I$ for a given field and define $I_C$ when the voltage exceeds $0.5~\mu$V (the extracted $I_C$ is not sensitive to this threshold). In Figure~\ref{Lambda}c, we plot $I_C(B)$ (circles) overlayed with fits to the theoretical predictions in the Meissner and mixed states~\cite{gaggioli_superconductivity_2024}, showing good agreement with the experimental results. The linear behaviour of the Meissner regime is clearly visible, before smoothly interpolating to the mixed regime when $B\approx 1.3$~mT. Similar behaviour of $I_C$ has been observed in thin metal films~\cite{plourde_influence_2001} and novel 2D superconductors with $W \ll \Lambda$~\cite{perego_experimental_2025}. 

An asymmetry in the surface barrier can result in a slight shift of the triangle-shaped current maximum away from zero $B$~\cite{gaggioli_superconductivity_2024}. This results in a superconducting diode effect at finite field, as recently demonstrated in KTaO$_3$-based devices~\cite{wang_ktao3-based_2025, yu_ktao3-based_2026} and other materials~\cite{hou_ubiquitous_2023}. We show this behavior in our samples in the appendix, Figure~\ref{supp_SDE}. With careful engineering of the surface barrier, KTaO$_3$ could act as a versatile platform for novel superconducting electronics when combined with the gate-tunability of the superconducting state~\cite{mccourt_electrostatic_2025}.

The slope of $I_C(B)$ in the Meissner regime allows one to extract the Pearl length $\Lambda$ from $\frac{dI_C}{dB}=-\frac{W^2}{\mu_0\Lambda}$~\cite{gaggioli_superconductivity_2024}, which we estimate as $\Lambda \approx1$~mm. While this equation is exact for a superconducting strip, we expect it to be correct, up to a geometric factor of order 1, in the case of a constriction of width $W$. From $\Lambda$, we directly calculate the superfluid stiffness, $J_s=\frac{\hbar^2}{2\mu_0 e^2\Lambda} \approx 15$~K, and kinetic inductance of $L_K =\frac{\hbar^2}{4e^2J_s}\approx 0.5$~nH (both at the base temperature). These quantities agree with the range found in other KTaO$_3$ samples~\cite{mallik_superfluid_2022, yu_sketched_2025, yang_tuning_2025}. We measure multiple $I_C(B)$ curves at different temperatures in the supplementary Figures~\ref{supp_WitchesHats} and~\ref{supp_IcIrVsT}. As $\Lambda\propto 1/n_s$ where $n_s$ is the Cooper pair density, we can plot the fraction of the superconducting electrons, $\rho(T)$ in Figure~\ref{Lambda}d. We fit it with the low-temperature expression for an isotropic s-wave superconductor at low temperatures~\cite{prozorov_magnetic_2006}, from which we find the gap at $T=0$ of $\Delta_0\approx180~\mu$eV, consistent with the transition temperature of about 1.3~K.

Next, in Figure~\ref{Lambda}e we compare the $I-V$  curves at $B=0$ and 5~mT, at three temperatures. All the curves show a similar behavior: starting at zero current, they abruptly switch from a low voltage state ($V<100~\mu$V) to the high voltage state ($V\sim 10$~mV) at the ``switching'' current. Due to the high dissipated power of several tens of nW, it is likely that in this regime a large portion of the 2DEG surrounding the constriction becomes normal. 

Upon reversal of the current sweep, the curves transition from the high to the low voltage state at the ``retrapping'' current. Note that while the switching currents strongly depend on the temperature and magnetic field, as expected from Figure~\ref{Lambda}c, the retrapping current is nearly independent of the temperature or the  field (in the few mT range). Furthermore, in the high resistance state the voltage also does not depend on $B$ and $T$, Figure~\ref{Lambda}e. This further supports the interpretation of the high-voltage state as the normal state sustained by Joule heating. 

We now zoom in by 3 orders of magnitude in voltage to the low voltage state. Figure~\ref{Lambda}f shows the same IV curves as in Figure~\ref{Lambda}e (upward ramp only), but on the $\mu$V scale rather than the mV scale. Here, we observe different behaviors at zero (or very small) $B$ vs larger values of $B$. Namely, at $B=0$, the curves switch from the noise floor (less than $1~\mu$V) straight to the high voltage state. At $B=5$~mT, there exists a low voltage regime with $V$ in the range of a few $\mu$V (this regime is visible in the map of Figure~\ref{Lambda}b as a red sliver at the transition between the black and white regions). We argue that this regime corresponds to flux flow, and return to it later in Figure~\ref{IVs}. 
For now, we turn our attention to the direct switch from zero resistance to the normal state at zero or very small magnetic field.

\section{\label{sec:histograms} Switching statistics and rates} 

\begin{figure*}[htp!]
\includegraphics[width=1\textwidth]{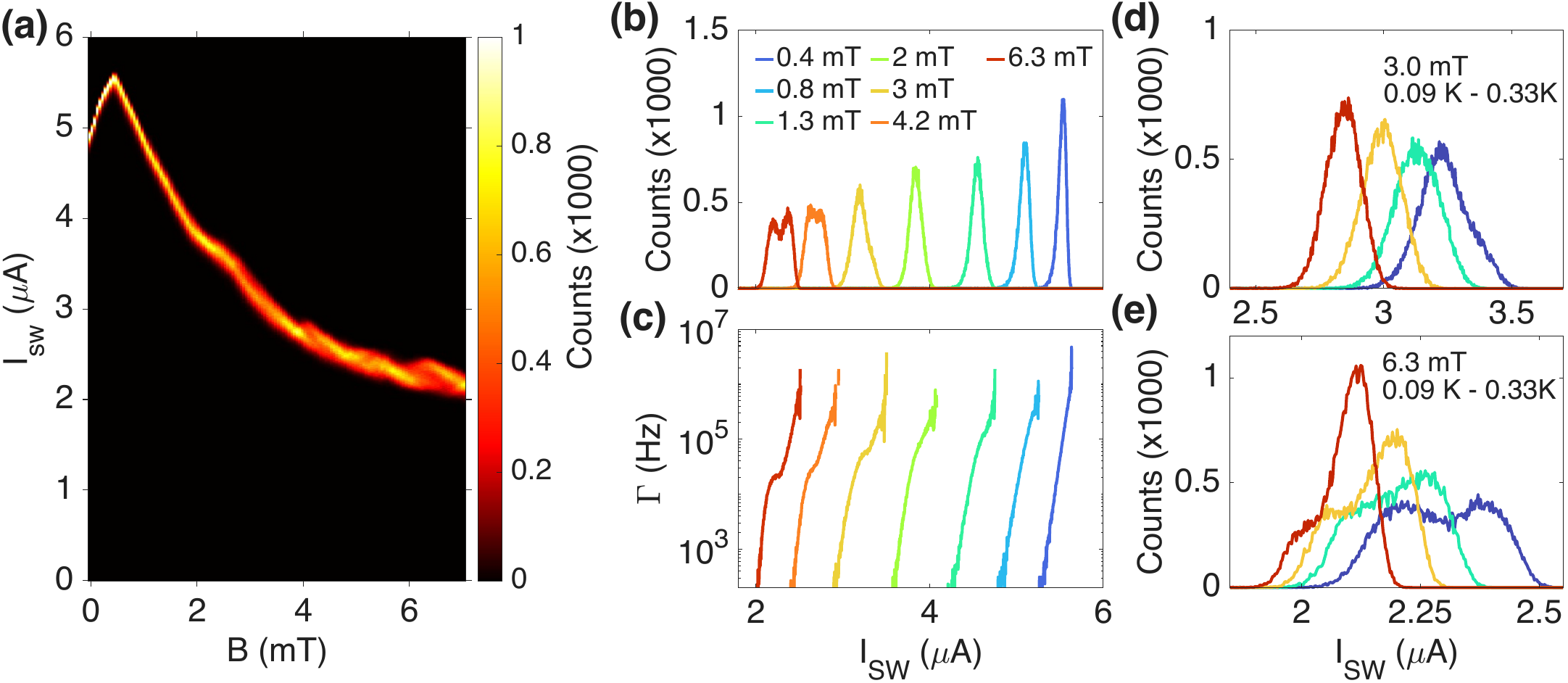}
\caption{\label{rate_B} \textbf{(a)} Heat map of multiple switching histograms like in Figure~\ref{rate_T}a, plotted as a function of $B$ at 0.09~K. \textbf{(b)} Histograms at various fields corresponding to the vertical cuts of panel (a). At $B \approx 1$~mT the histograms lose their asymmetric shape before transitioning to distributions with multiple peaks. \textbf{(c)} Escape rates calculated from histograms in panel (b) under the assumption of a single population of the initial states. \textbf{(d,e)} Histograms of switching currents at 3.0~mT (top) and 6.3~mT (bottom) for 0.09~K (blue), 0.24~K, 0.29~K and 0.33~K (red).}
\end{figure*}

\begin{figure*}[htp!]
\includegraphics[width=0.8\textwidth]{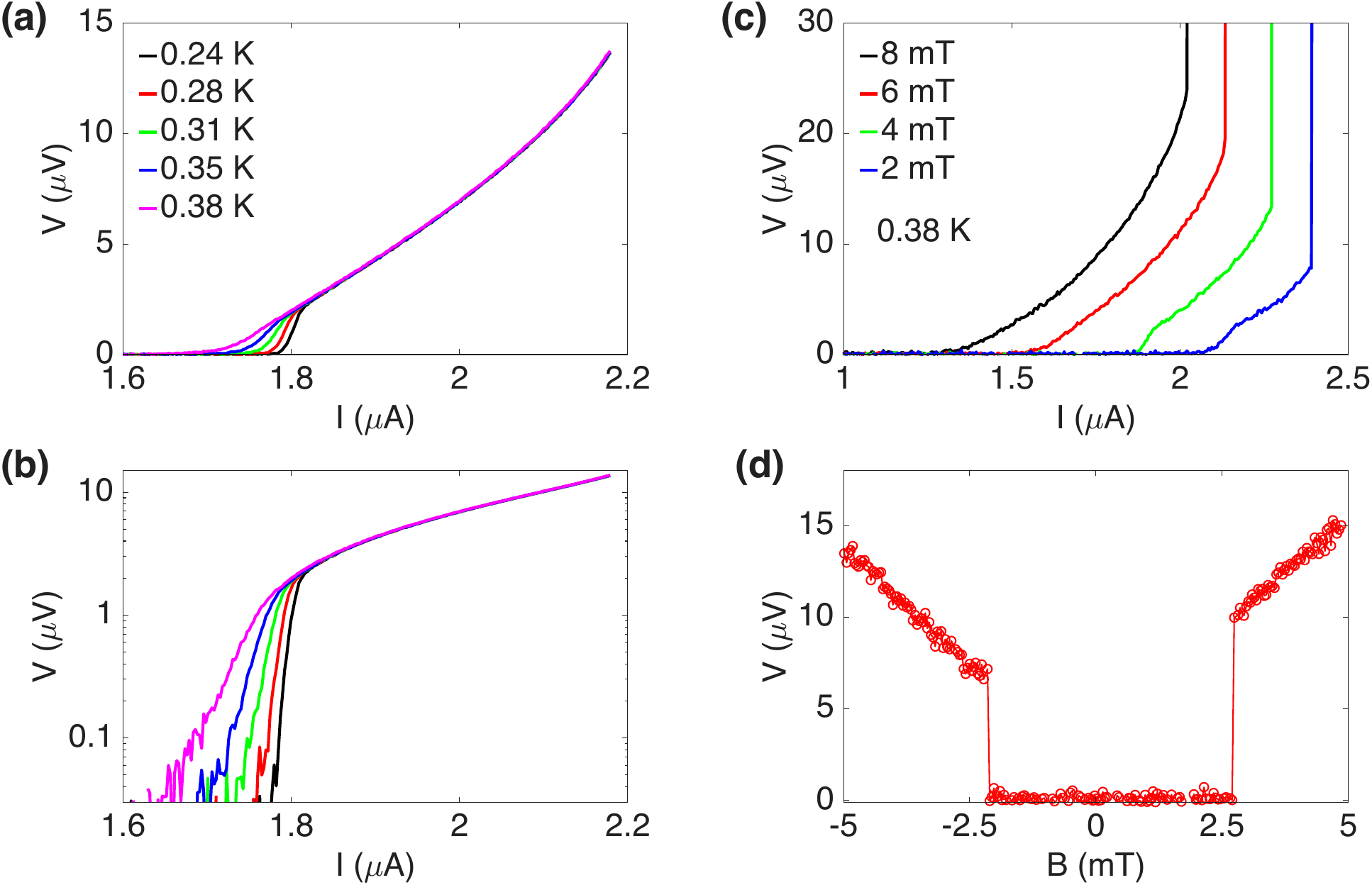}
\caption{\label{IVs} \textbf{(a)} $I-V$ curves in the $\mu$V regime  measured in device A for a range of temperatures $T \gtrsim 0.24$K at $B=5.6$~mT. Note that the curves are temperature-independent for $I \gtrsim 1.82 \mu$A. \textbf{(b)} The same data from (a) plotted on a logarithmic voltage scale, demonstrating the exponential temperature-dependent tail. \textbf{(c)} $I-V$ curves at several values of $B$ showing the vortex flow regime and the jump to the normal state. The voltage at the end point of the vortex flow regime is extracted for the next panel. \textbf{(d)} Voltage measured immediately prior to switching from the vortex flow ($\mu$V) regime to the normal (mV) regime, plotted as a function of magnetic field.}
\end{figure*}

We are now interested in exploring the nucleation of vortices at the edge of the constriction. We first present measurements at zero field (Figure~\ref{rate_T}), so that the pinned vortices should be absent. Here, and in Figure~\ref{rate_B}, we work with device B, nominally identical to device A studied in Figures~\ref{Lambda} and~\ref{IVs}.

Upon multiple sweeps of the $I-V$ curves, the transition from the zero to mV scale voltage occurs at slightly different values of bias current. We collect the statistics of the switching events from multiple linear current sweeps~\cite{fulton_lifetime_1974}. Following the nucleation of the vortex, the $I-V$ curve switches to the normal state. For each sweep, the value of the switching current $I_S$ is recorded, and the resulting histograms of $I_S$ are shown in Figure~\ref{rate_T}a. Clearly, the average $I_S$ increases as temperature decreases, before the distribution saturates at the lowest temperatures. Using established methods~\cite{fulton_lifetime_1974}, we can convert the switching histogram to the average switching rate, $\Gamma$, plotted in Figure~\ref{rate_T}b with the same horizontal axis. 

In Figure~\ref{rate_T}c, we plot the standard deviation (i.e. the width) of the distribution, and the maximal switching current, identified by the sharp drop-off of the histograms at their upper end. The three regimes familiar from the Josephson junctions are clearly visible: the phase diffusion ($T\gtrsim 0.45$ K), the thermally activated switching with the histogram width growing with temperature (0.2 K $\lesssim T\lesssim 0.4$K)~\cite{kurkijarvi_intrinsic_1972}, and the low-temperature saturation ($T\lesssim 0.15$ K). 

We are particularly interested in the third regime, potentially corresponding to the macroscopic quantum tunneling of vortices from the nucleation site at the edge of the constriction. The quantum tunneling of a vortex is fundamentally distinct from the macroscopic quantum tunneling observed in underdamped Josephson junctions~\cite{clarke_quantum_1988}. While both phenomena can be formulated as tunneling of a particle coupled to dissipative environment, vortex tunneling corresponds to the real-space motion of a topological excitation across a barrier. Furthermore, the motion of the vortex may be expected to be overdamped. Signs of the macroscopic quantum tunneling of vortices have been reported in the $I-V$ curves of High $T_C$ films~\cite{tafuri_dissipation_2006}, for  Josephson vortices in annular junctions~\cite{wallraff_quantum_2003}, 
and very recently based on the vortex entry and exit rates in magic angle twisted graphene~\cite{perego_pearl-vortex_2026}.

We expect the rate of this process to scale as $\Gamma \propto (1-i) \exp(-\alpha(1-i))$, where $\alpha$ is a constant, $i=I/I_C$ and $I_C$ is the ``true critical current" at which the energy barrier for vortex nucleation vanishes~\cite{weiss_quantum_2021}. The expression above corresponds to tunneling with strong dissipation, expected in the case of vortices and similar to the behavior of a Josephson junction in the overdamped regime. The slight deviation of the switching rate from the linear slope on the log-linear plot in Figure~\ref{rate_T}b can be adequately accounted for by the prefactor $1-i$. In Figure~\ref{rate_T}d, we plot $\Gamma/(1-i)$ with $I_C=5.05~\mu$A, which clearly works well.

Next, we explore the switching process as a function of the magnetic field. Figure~\ref{rate_B}a shows the map made of multiple switching histograms as a function of the magnetic field. The $\mu$V-scale flux-flow regime we have encountered in Figure~\ref{Lambda}f does not set in until $B\sim$~12~mT in this constriction (supplementary Figure~\ref{supp_QPC3Fraunhofer}) and all the switching events in Figure~\ref{rate_B}a take the system from zero to the normal state. The individual histograms corresponding to select vertical cuts of the map are shown in Figure~\ref{rate_B}b, and the calculated switching rates in Figure~\ref{rate_B}c. The measurement shown in Figure~\ref{rate_T} corresponds to the $B=0$ curve (dark blue). 

As we move away from $B=0.4$~mT, which corresponds to the highest critical current, we find the curves deviate further and further from the typical Gumbel distribution seen in Figure~\ref{rate_T}a: they become less asymmetric and eventually develop sub-peaks. The corresponding calculated escape rates demonstrate two offset segments with a similar slope (see the red curve at $B=6.3$~mT). Figure~\ref{rate_B}a shows that these peak doublings correspond to transitions between different branches on the map, with the lower current branches waning as the field increases, while being replaced by the upper branches. 

We argue that these branches correspond to the different vortex configurations -- the next branch adds one vortex (or more) to the constriction, which reduces the diamagnetic current at the edge, thereby increasing the critical current. The presence of the two peaks in the histogram corresponds to two distinct populations: for a given sweep, the number of vortices in the constriction is likely fixed following the cooling off following the previous sweep. We note that in the case when two sub-populations are present, one cannot rely on the values of the switching rate in Figure~\ref{rate_B}c, extracted using the expressions in Ref.~\cite{fulton_lifetime_1974}, which assumes a single population. As the temperature increases, the finite-field histograms may begin to recover some of the asymmetry (Figure~\ref{rate_B}d), or, in some cases, one peak may come to dominate the other (Figure~\ref{rate_B}e). This behavior suggests that the distinct populations contributing to the measured histograms evolve on different temperature scales.

\section{\label{sec:IV}I-V curves: vortex flow and creep}

We now return to the  sample shown in Figure~\ref{Lambda} and discuss the $I-V$ curves at finite field in more detail.

First, we plot the low-voltage regime for $B=5.6$~mT at several temperatures in Figure \ref{IVs}a. The same curves are plotted on a logarithmic voltage scale in Figure~\ref{IVs}b. 
We observe that the lower end of the IV curves (below $I=1.8~\mu$A) appears linear on the log-linear plot, indicating an exponential dependence of the voltage on bias current, with a slope that depends on  temperature. In contrast, the higher end of this low-voltage regime (above $\sim 1.82~\mu$A) does not depend on temperature, and here multiple curves coalesce. 
Clearly, this is not the normal state, because the voltage here is $\sim 1000$ times lower than that in the high-voltage state of Figure \ref{Lambda}e. 

We tentatively attribute this temperature-independent regime (above $\sim 1.82~\mu$A) to the vortex flow, in which the vortices are continuously crossing the constriction. Given the dimensions of the constriction and the magnitude of the field, we estimate that the number of vortices present in the constriction at a given time is of the order of one. Each time a vortex crosses the constriction results in a voltage pulse of $\int V dt=h/2e$. The measured voltage is equal to the average of the pulses, which is therefore proportional to the average velocity of the vortices. This consideration allows us to estimate that on average a vortex crosses the constriction once every $0.2$~ns, corresponding to the velocity of $5$~km/sec if only one vortex is present at a given time. This scale, and even higher velocities have been previously reported~\cite{embon_imaging_2017}.

We now turn to the lower end of the curves in Figure~\ref{IVs}a and Figure~\ref{IVs}b (below $I=1.8~\mu$A). We speculate that the voltage in this regime is limited by the rate at which vortices are either nucleated at the edge of the constriction or depinned in the bulk. Once they start moving, the vortices cross the constriction relatively fast, so that the measured voltage is proportional to the rate of the depinning or nucleation events, which in turn depends on temperature. Notably, these rates are suppressed at low temperature, and below $T \sim 200$~mK, the exponential tail of the $I-V$ curves disappears being replaced by stochastic switching from zero straight to the flux flow regime (supplementary Figure~\ref{IsIr_B}).  

The vortex flow regime is bound on the upper end by an abrupt switching to the normal state. In Figure~\ref{IVs}c, we show several such jumps for four values of magnetic field. Unlike the switching from the zero-voltage state, this jump occurs at the same value of current upon multiple current sweeps -- see the dark blue and magenta curves in Figure~\ref{Lambda}f, measured at $T=0.07$ and 0.57~K. 

We speculate that at the end point of the vortex flow regime, an instability develops because the viscous drag force acting on the vortices reaches its maximal possible value. (See Ref.~\cite{dobrovolskiy_fast_2024} for a review of the possible mechanisms.)  The voltage at that point is found to be proportional to the applied magnetic field, as shown in Figure~\ref{IVs}d. 

The critical current for this instability is only weekly dependent on magnetic field, see the red-white boundary in Figure~\ref{Lambda}b. Since at small $B$ the vortex nucleation threshold occurs at higher currents, the flux flow regime disappears below a certain field, and the system switches directly from the zero voltage to the normal state. This occurs for $|B|\lesssim 2$~mT for the first sample (Figure~\ref{Lambda}b) and $B\lesssim 10$~mT for the second sample (Figure~\ref{supp_QPC3Fraunhofer}a). 

\section{\label{sec:conclusion}Conclusion}
In this paper, we identify and explore several distinct regimes: 

In the mixed state at finite field, we observe the regime in which the $I-V$ curves are (nearly) independent of temperature but strongly depend on the magnetic field. We argue that in this regime a few, and possibly just one vortex continuously shuttles across the constriction. 

For smaller values of current, and elevated temperatures ($T\gtrsim 250$~mK) the voltage is found to depend exponentially on the current, and the slope of $\log V$ vs $I$ depends on temperature. Here, the rate of vortex crossing the constriction is limited by their nucleation at the edge or depinning from the bulk. 

Finally, we study the nucleation of the vortices at zero magnetic field as a function of applied bias. We measure the dependence of the escape rate on the proximity to the critical current $I_C$, which follows the semiclassical dependence for tunneling with a strong dissipation.

Overall, we find that the superconducting state at the interface of AlO$_x$ and KTaO$_3$ represents a very promising playground for exploring the vortex dynamics in a 2D superconductor.\\
\\

\section{\label{sec:Acknowledgments}acknowledgments}
\begin{acknowledgments}
\noindent We appreciate fruitful discussions with Filippo Gaggioli. J.T.M. thanks Stephen Teitsworth for insightful discussions regarding tunneling scaling.  \\
\\
Sample fabrication by R.H. and J.T.M., transport measurements by J.T.M., J.C. and C.C. and data analysis by J.T.M. and G.F. were supported by the National Science Foundation grant DMR-2327535. Crystal synthesis by D.K. and S. were supported by the U.S. National Science Foundation under Grant No. NSF DMR-2324174. Sample fabrication by R.H. and J.T.M. was performed in part at the Duke University Shared Materials Instrumentation Facility (SMIF), a member of the North Carolina Research Triangle Nanotechnology Network (RTNN), which is supported by the National Science Foundation (award number ECCS-2025064) as part of the National Nanotechnology Coordinated Infrastructure (NNCI).

\end{acknowledgments}

\bibliography{CritState_Refs}

\appendix
\section{Appendix}

\subsection{Time-reversal symmetry}

Figure~\ref{supp_FraunhoferUpAndDown} compares the voltage maps of device A for positive and negative current sweep directions. We present the second map flipping the direction of the current and field axes. Clearly, reversing the sign of $I$, $V$, and $B$ results in very similar maps. This indicates that any intrinsic magnetism associated with d-orbitals of the Ta ions is negligible in this sample. 

\begin{figure*}[htp!]
\includegraphics[width=0.9\textwidth]{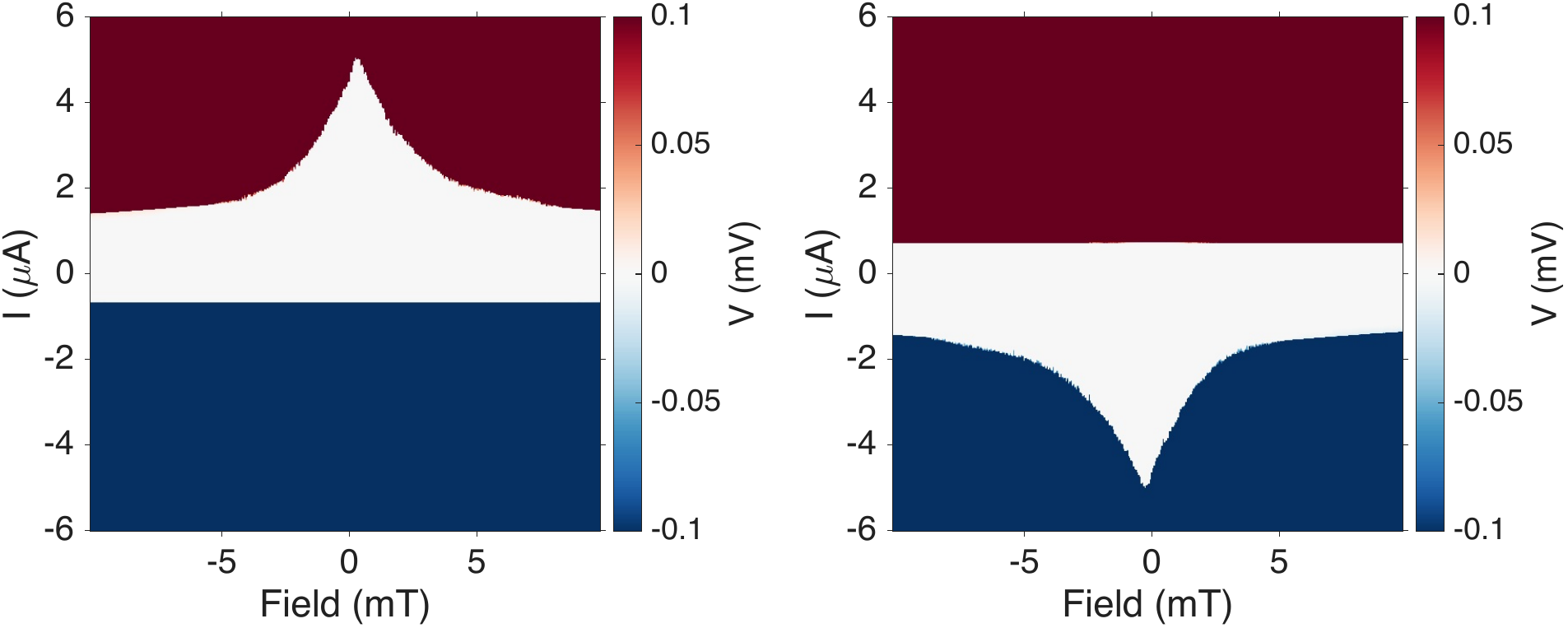}
\caption{\label{supp_QPC3UpAndDown} Bias-field maps of measured voltage in the device B, as presented in Figure~\ref{rate_T} and Figure~\ref{rate_B}. The left plot displays the voltage on an upwards sweep on a logarithmic scale. Upwards and negative sweeps of bias are shown in the middle and right plots respectively.}
\end{figure*}

Figure~\ref{supp_QPC3UpAndDown} shows similar maps for device B. Here, the axes are not inverted, and the time-reversal symmetry is demonstrated by the 180$^{\circ}$ rotation symmetry of the two maps measured for the up and down current sweep directions. Because of the asymmetry in the device, it can be seen how a diode effect may occur at finite field.

\subsection{Superconducting Diode Effect}

\begin{figure*}[htp!]
\includegraphics[width=0.9\textwidth]{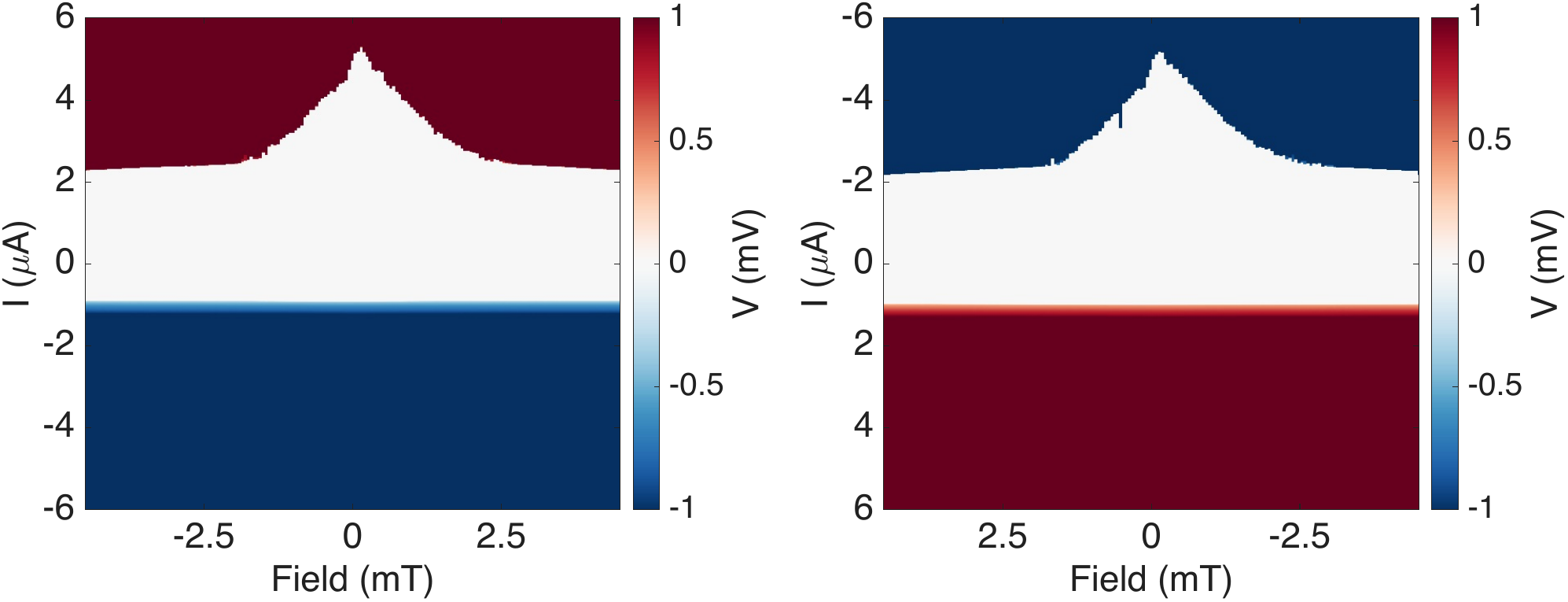}
\caption{\label{supp_FraunhoferUpAndDown} Measured voltages in the device presented in Figures~\ref{Lambda} and~\ref{IVs} at 0.08~K. The left (right) plot displays the voltage as the current is swept from negative (positive) to positive (negative). On the right plot, the x and y axis have both been flipped.}
\end{figure*}

\begin{figure*}[htp!]
\includegraphics[width=0.8\textwidth]{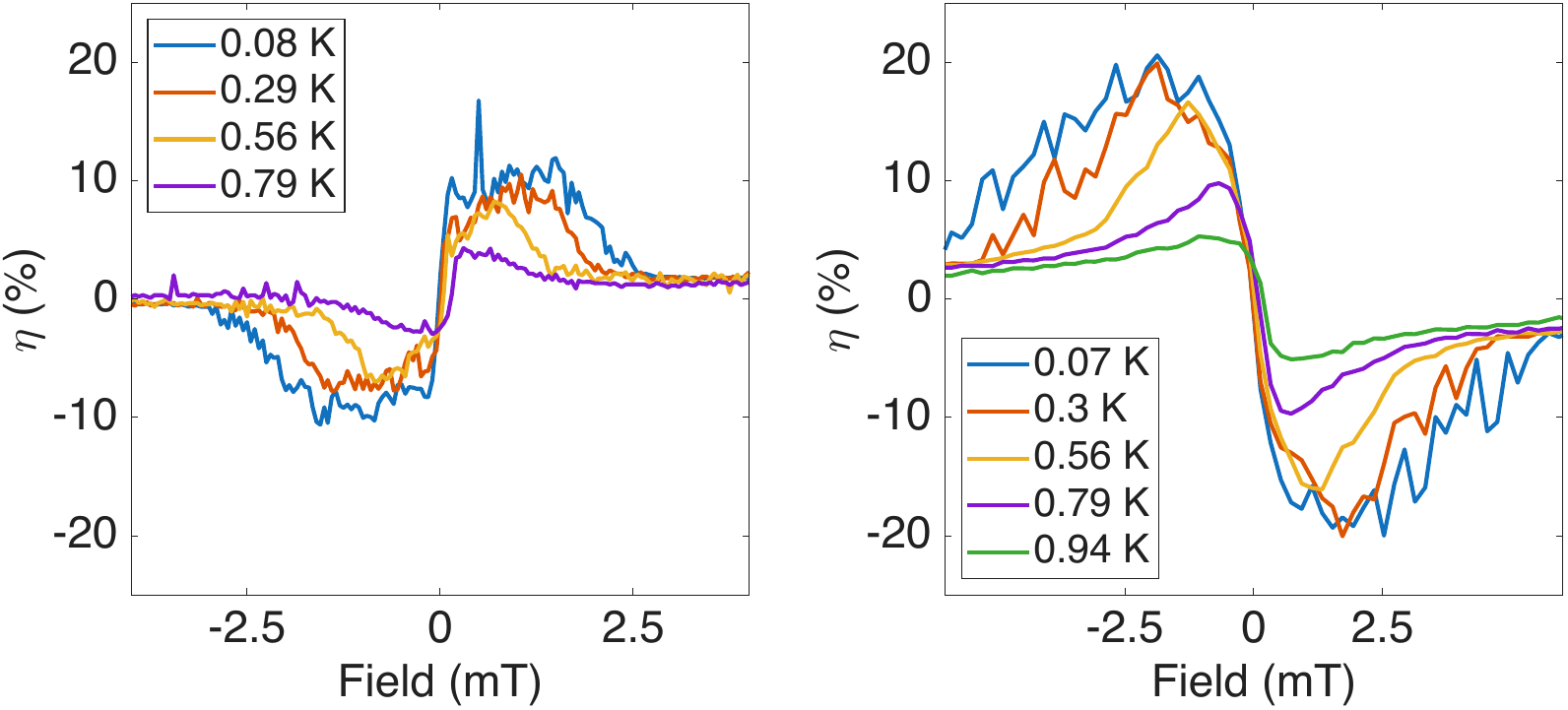}
\caption{\label{supp_SDE} Superconducting diode efficiency ($\eta$) extracted directly from Figure~\ref{supp_FraunhoferUpAndDown} and at various temperatures (left). The right plot displays another device with a greater surface barrier asymmetry, and as a result a larger diode efficiency.}
\end{figure*}

In Figure~\ref{supp_SDE}, we display  the superconducting diode efficiency ($\eta$) extracted directly from Figure~\ref{supp_FraunhoferUpAndDown} and at various temperatures (left). The efficiency is defined as $\eta=\frac{I_{C+} - |I_{C-}|}{I_{C+} + |I_{C-}|}$, where $I_{C+}$ and $I_{C-}$ are the critical current at positive and negative bias sweep respectively. The right plot displays $\eta$  for the third constriction (C) made simultaneously with device A. It appears to have a greater surface barrier asymmetry, and as a result a larger diode efficiency. The diode effect decreases with temperature but still persists even close to $T_C$.

\subsection{Bias-field Characteristics at Various Temperatures}

\begin{figure*}[htp!]
\includegraphics[width=\textwidth]{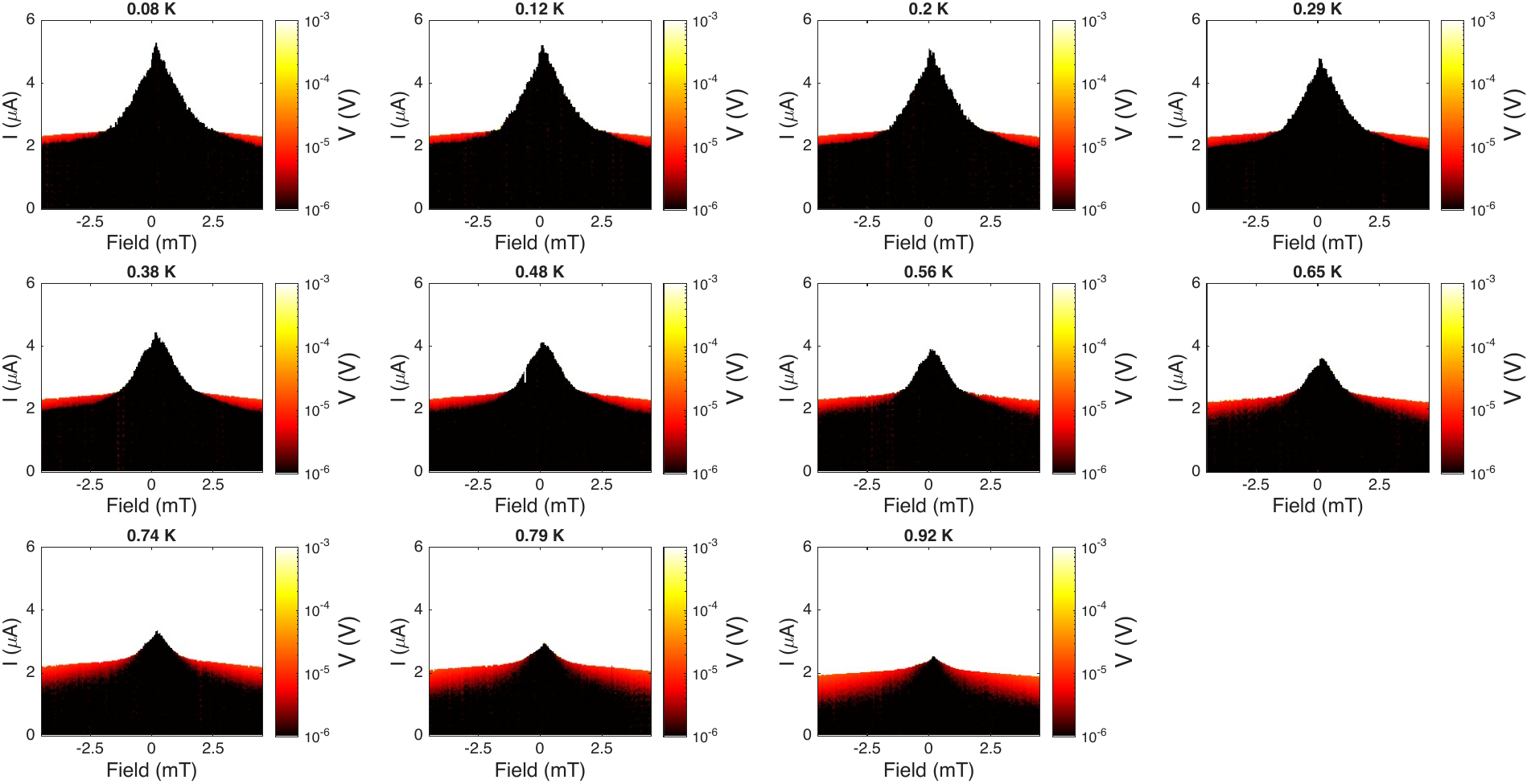}
\caption{\label{supp_WitchesHats} Measured voltages in the device presented in Figures~\ref{Lambda} and~\ref{IVs} at various temperatures.}
\end{figure*}

Figure~\ref{supp_WitchesHats} shows voltage maps of device A at different temperatures from 0.08~K to 0.92~K. The first panel is identical to Figure~\ref{Lambda}b. 
The critical state model of Ref.~\cite{gaggioli_superconductivity_2024} appears to qualitatively describe the  $I_C$ behaviour up to higher temperatures very well. The range of the Meissner regime decreases with increasing temperature. As the range of the Meissner regime shrinks in field, the vortex flow regime (red) beings to appear at lower field but never at zero field. Using these measurements, we extract $\Lambda$ at various temperatures, as presented in Figure~\ref{Lambda}d.

\subsection{$I_C$ and $I_R$ vs temperature}

\begin{figure*}[htp!]
\includegraphics[width=0.6\textwidth]{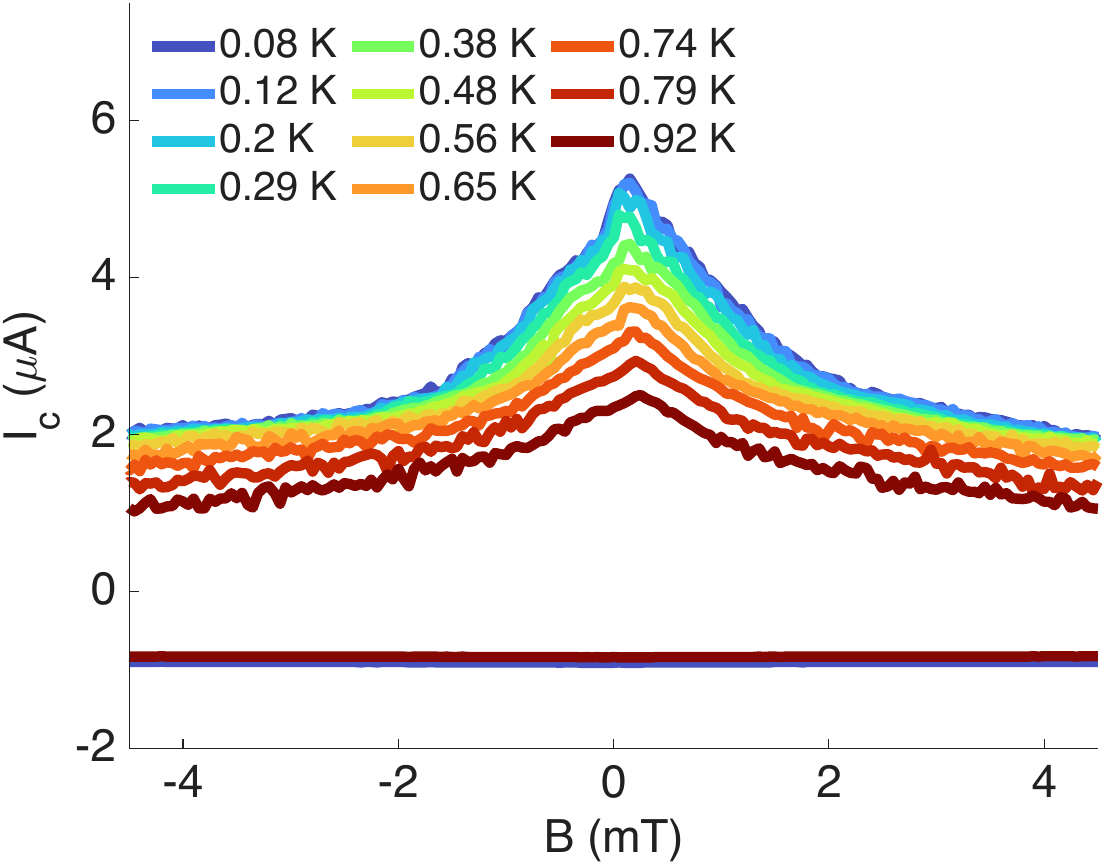}
\caption{\label{supp_IcIrVsT} Critical current and retrapping current at various temperatures in the device A.}
\end{figure*}

It can be seen from Figure~\ref{supp_IcIrVsT} that the critical current $I_C$ changes significantly from 0.08~K to 0.92~K, however in the same range of temperatures, the retrapping current $I_R$ changes very little. This suggest that whilst the device is undergoing dissipation at the mV scale, the local temperature is significantly higher than the measured temperature.

\subsection{Retrapping in the flux flow regime}

\begin{figure*}[htp!]
\includegraphics[width=0.9\textwidth]{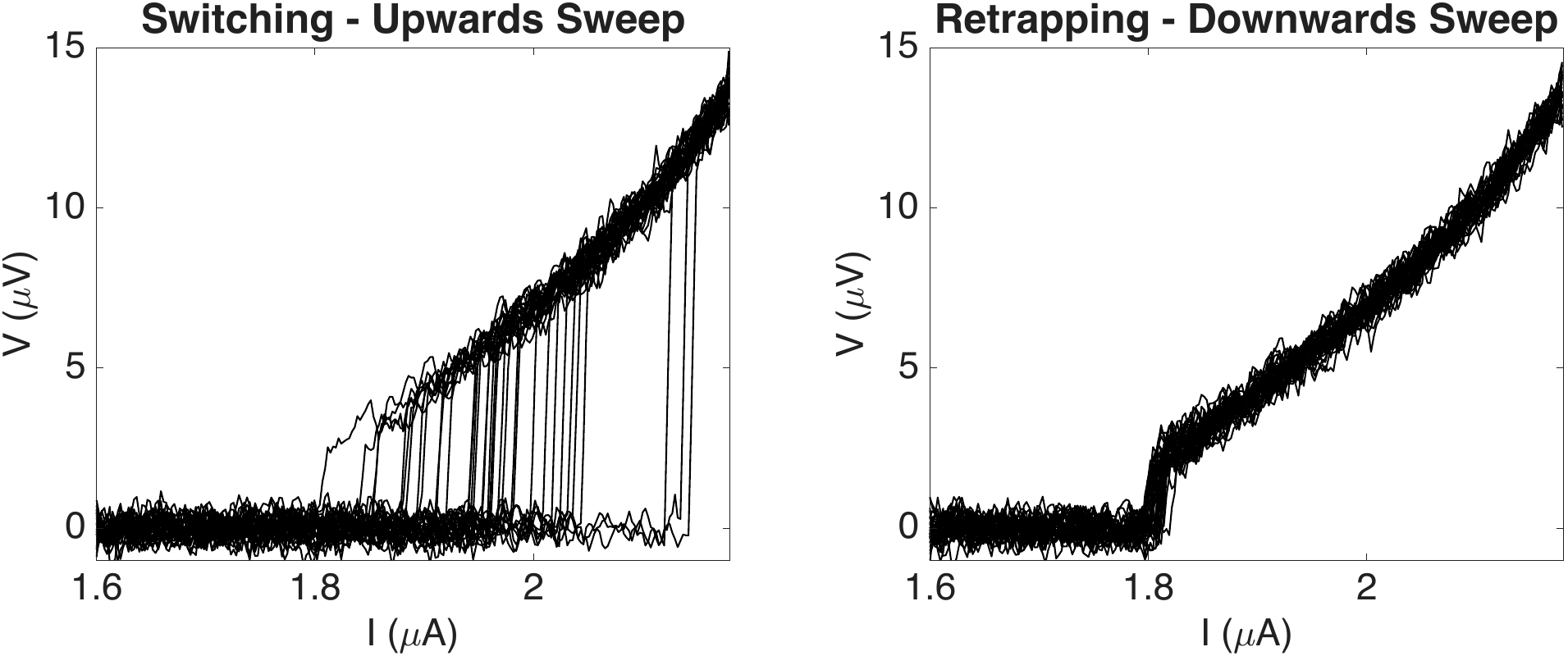}
\caption{\label{IsIr_B} Upwards and downwards sweeps of bias at 5.6~mT and 0.07K, repeated 35 times.}
\end{figure*}

The curves in Figure~\ref{IVs}a,b are measured at elevated temperatures, $T>240$~mK. At lower temperatures, the nucleation or depinning rate of vortices is suppressed, and the low-voltage curves become stochastic, as shown in Figure~\ref{IsIr_B}a. Here, we show 35 upward sweeps from $1.6\mu A$ to 2.2~$\mu A$. Upon successive sweeps, the voltage jumps from the sub-$\mu$V level to the few-$\mu$V range at different values of the current. Upon ramping the current down, the retrapping occurs at a lower value of current, Figure~\ref{IsIr_B}b. Note that this switching and retrapping behavior is similar to that shown in Figure~\ref{Lambda}e,f. The difference is that here the switching brings the system to the vortex flow regime rather than the normal state of Figure~\ref{IVs}a.

\subsection{Zero-field Switching Current Heatmap}

\begin{figure*}[htp!]
\includegraphics[width=0.6\textwidth]{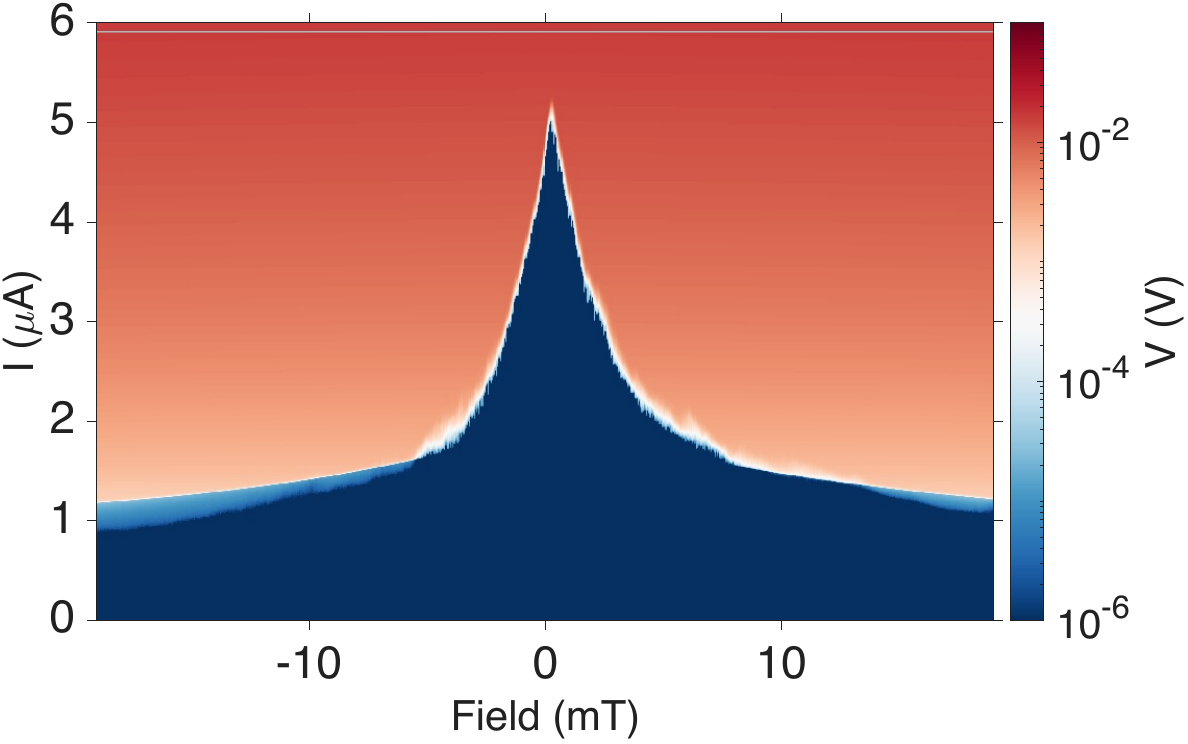}
\caption{\label{supp_QPC3Fraunhofer} Bias-field map of voltage in device B, plotted on a logarithmic colour scale.}
\end{figure*}

In Figure \ref{supp_QPC3Fraunhofer} we plot a bias-field map of the voltage in device B, plotted on a logarithmic colour scale and for an expanded range of field. It can be seen from the map that the vortex flow regime (light blue) does not appear until at positive field until $12$~mT.

\subsection{Zero-field Switching Current Heatmap}

\begin{figure*}[htp!]
\includegraphics[width=0.6\textwidth]{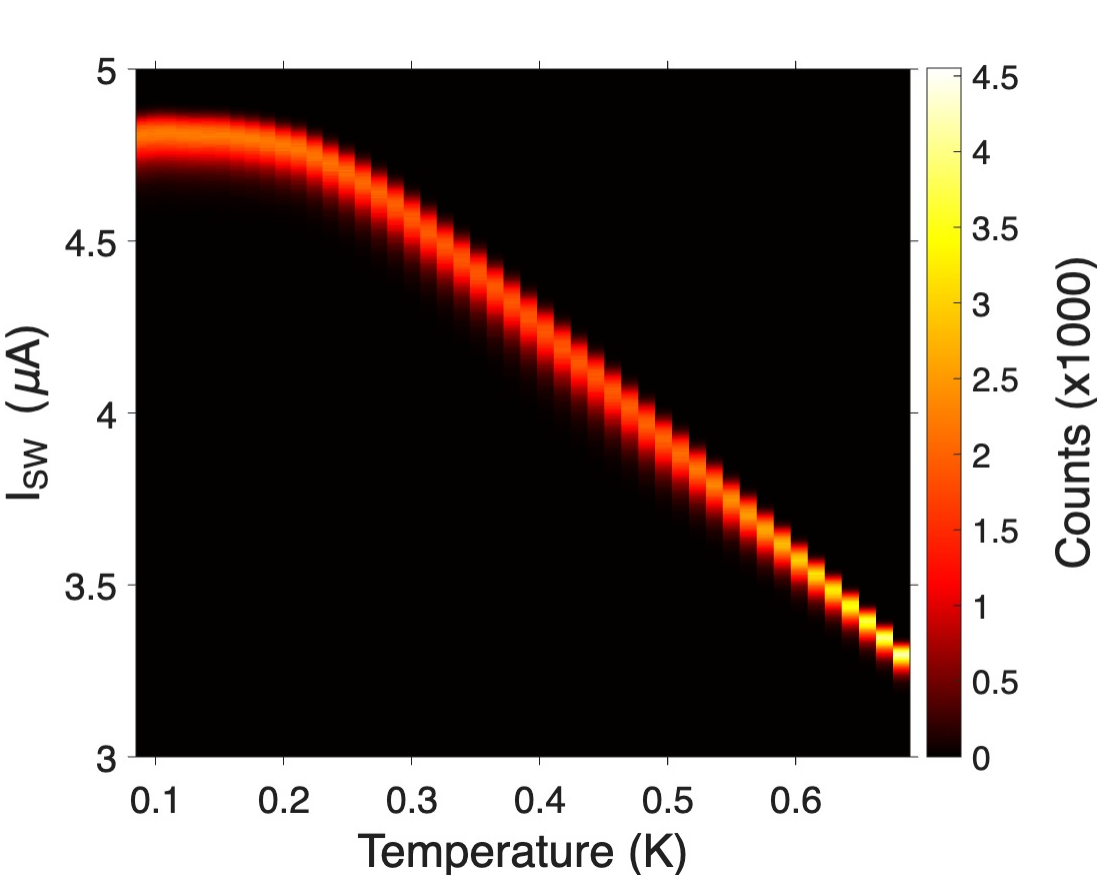}
\caption{\label{supp_Fig3Heatmap} Heatmap of switching currents at zero field, corresponding to measurements presented in Figure~\ref{rate_T}.}
\end{figure*}

Measurements of the distribution of switching currents at zero-field and a range of temperatures is presented in Figure~\ref{supp_Fig3Heatmap}. The histograms presented in Figure~\ref{rate_T} correspond to cuts in Figure~\ref{supp_Fig3Heatmap}. The distribution remains unimodal across all temperatures, with the mean decreasing with increasing temperature. 

\subsection{Zero-field Switching IV-Curves}

\begin{figure*}[htp!]
\includegraphics[width=0.6\textwidth]{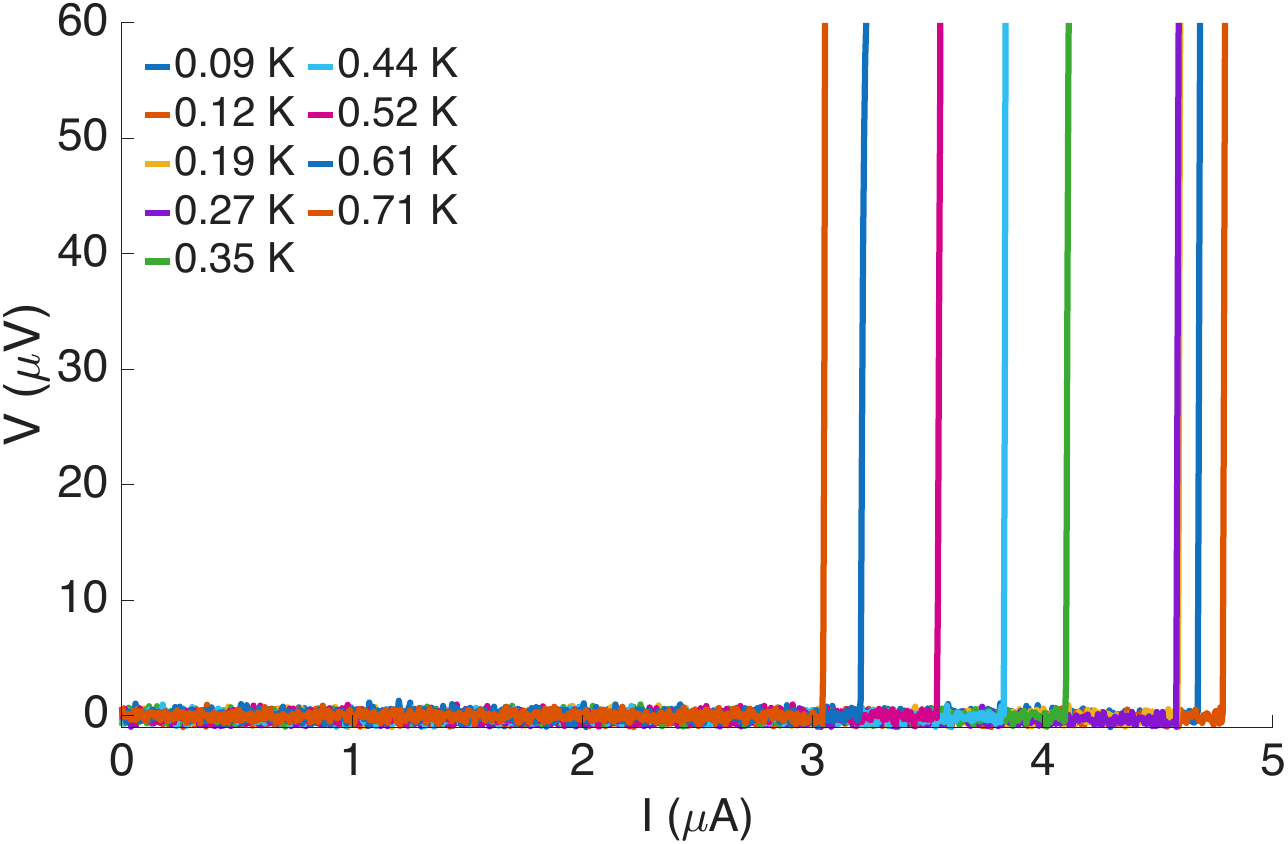}
\caption{\label{supp_ZeroFieldIVs} IV curves displaying individual switching events of the device presented in Figure~\ref{rate_T} and Figure~\ref{rate_B}}
\end{figure*}

IV curves displaying individual switching events of the device B are presented in Figure~\ref{supp_ZeroFieldIVs} at a range of temperatures. It can be seen that the $\mu$V scale dissipation, which we associate with vortex flow, is not present even at higher temperatures, up to atleast 0.71~K.

\subsection{Switching Current Heatmap at 6.3~mT}

\begin{figure*}[htp!]
\includegraphics[width=0.9\textwidth]{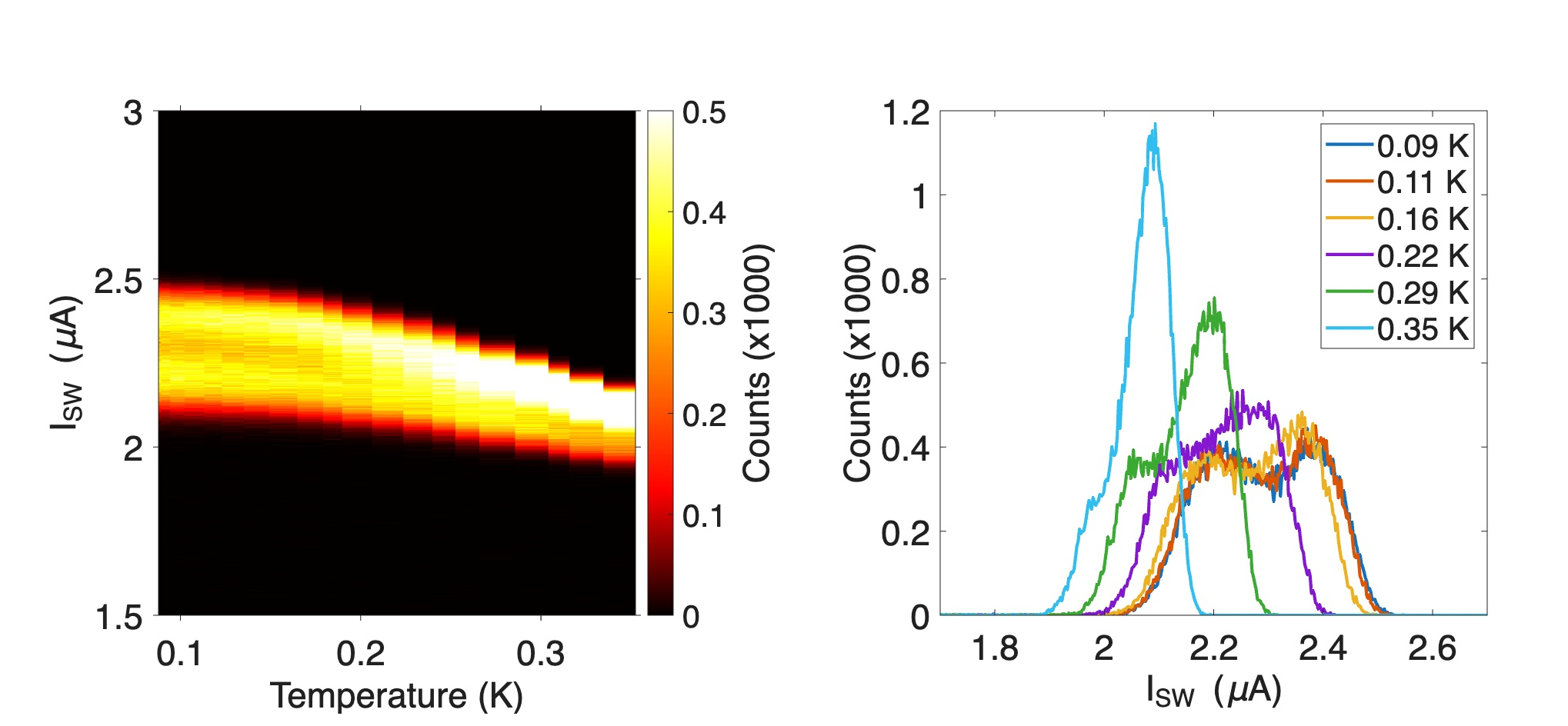}
\caption{\label{supp_DoublePeakHeatmap} Heatmap of switching currents at 6.3~mT, corresponding to measurements presented in Figure~\ref{rate_B}. Cuts at various temperatures are also presented (right). Note that the colourbar has been capped in order to accentuate the double-peak structure at low temperatures.}
\end{figure*}

Measurements of the distribution of switching currents at 6.3~mT and a range of temperatures is presented in Figure~\ref{supp_DoublePeakHeatmap}. It can be seen that the double-peak structure evolves non-trivially with temperature. As the temperature increases, the peaks appear to merge, resulting in one dominant peak. If we interpret the histogram as two independent distributions summed, it could also be the case that one of these distributions dominates over the other. In this case the distribution with a larger mean $I_{SW}$ dominates at higher temperatures. Note that the temperature steps are uneven, meaning that some of the histograms appear elongated in the x-direction.

\end{document}